\documentclass[12pt,a4paper]{article}
\usepackage{color,graphics,amsmath,epsfig,rotating}

\begin{document}

\setlength{\unitlength}{1mm}
\renewcommand{\arraystretch}{1.4}


\def\micromegas      {{\tt micrOMEGAs}}
\def\cpsh            {{\tt CPsuperH}}

\def\mxs{M_{X_S}}
\def\cnur{c_{\nu_R}}
\def\cnul{c_{\nu_L}}
\def\ctl{c_{t_L}}
\def\ctr{c_{t_R}}
\def\mzp{M_{Z'}}
\def\mlzp{M_{LZP}}
\def\mneut{M_{\tilde{\chi}^0_1}}
\def\sw{s_W}
\def\cw{c_W}
\def\ca{\cos\alpha}
\def\cb{\cos\beta}
\def\sa{\sin\alpha}
\def\sb{\sin\beta}
\def\ssi{\sigma^{SI}_{\chi N}}
\def\si{\sigma^{SI}}
\def\sip{\sigma^{SI}_{\chi p}}
\def\ssd{\sigma^{SD}_{\chi N}}
\def\sd{\sigma^{SD}}
\def\sdp{\sigma^{SD}_{\chi p}}
\def\sdn{\sigma^{SD}_{\chi n}}
\def\micro{micrOMEGAs}

\newcommand{\beqn}{\begin{eqnarray}}
\newcommand{\eeqn}{\end{eqnarray}}

\newcommand{\eq}[1]  {\mbox{(\ref{eq:#1})}}
\newcommand{\fig}[1] {Fig.~\ref{fig:#1}}
\newcommand{\Fig}[1] {Figure~\ref{fig:#1}}
\newcommand{\tab}[1] {Table~\ref{tab:#1}}
\newcommand{\Tab}[1] {Table~\ref{tab:#1}}

\newcommand{\gsim}{\;\raisebox{-0.9ex}
           {$\textstyle\stackrel{\textstyle >}{\sim}$}\;}

\newcommand{\lsim}{\;\raisebox{-0.9ex}{$\textstyle\stackrel{\textstyle<}
           {\sim}$}\;}

\newcommand{\smaf}[2] {{\textstyle \frac{#1}{#2} }}
\newcommand{\sfrac}[2] {{\textstyle \frac{#1}{#2}}}
\newcommand{\pif}      {\smaf{\pi}{2}}

\def\delr            {\!\stackrel{\leftrightarrow}{\partial^\mu}\!}

\newcommand{\bb} {\color{blue}}
\newcommand{\bbf} {\color{blue}\bf}
\newcommand{\change} {\bf\mbf}


\begin{flushright}
   \vspace*{-18mm}
   Date: \today
\end{flushright}
\vspace*{2mm}

\begin{center}

{\Large\bf Discriminating dark matter candidates  using direct detection} \\[8mm]

{\large   G.~B\'elanger$^1$, E. Nezri$^2$,  A.~Pukhov$^3$}\\[4mm]

{\it 1) LAPTH, Univ. de Savoie, CNRS, B.P.110,  F-74941 Annecy-le-Vieux, France\\
     2) LAM,CNRS, Univ. Aix-Marseille I, 2 place le Verrier, 13248 Marseille, France\\
     3) Skobeltsyn Inst. of Nuclear Physics, Moscow State Univ., Moscow 119992, Russia  }\\[4mm]

\end{center}

\begin{abstract}
We examine the predictions for both the spin dependent and spin independent direct detection rates in a variety of 
new particle physics models with dark matter candidates. We show that a determination of both  spin independent and spin dependent amplitudes
 on protons and neutrons can in principle 
discriminate different candidates of dark matter up to a few ambiguities. We emphasize the
importance of making measurements with different  spin dependent sensitive detector materials and the need for significant improvement of the 
detector sensitivities.  Scenarii where exchange of  new coloured particles 
contributes  
significantly to the elastic scattering cross sections are often the 
most difficult  to identify, the LHC should give an indication whether such scenarii are relevant for direct detection.
\end{abstract}

\section{Introduction}

 Unraveling
 the properties of a new stable cold dark matter (CDM) particle is a challenge for ongoing or future 
 astroparticle and collider experiments.
The most convincing evidence for CDM so far is provided  by WMAP~\cite{Spergel:2006hy} 
and SDSS~\cite{Tegmark:2006az}. 
Their  precise determination 
of the relic density of CDM strongly constrains the parameter space of the various new particle physics
models(NP)~\cite{Ellis:2003cw,Profumo:2004at,Chattopadhyay:2003xi,Baer:2003yh,Hooper:2007qk,Kong:2005hn,Burnell:2005hm,Servant:2002aq}. 
This single observable, $\Omega_{CDM}h^2$,  is however not sufficient to pin down the properties of CDM even when
assuming that the candidate is a weakly interacting massive particle, $\chi$. 
Additional information on the nature of dark matter could  also  be obtained 
from measurements of
detection rate in different detectors,
observations  of a signal in photons, antiprotons, positrons or
neutrinos produced after annihilation  of dark matter
and discovery and measurements of properties of new particles at colliders.

Several models for new physics  containing a CDM candidate have been proposed in 
the past~\cite{Bertone:2004pz}. The most  popular examples of new  stable weakly interacting particles
at the electroweak scale include 
the neutralino in supersymmetric models~\cite{Goldberg:1983nd,Ellis:1983ew}, right-handed neutrinos~\cite{Agashe:2004ci,Hsieh:2006qe,Asaka:2006fs}, scalars or vector bosons in
extra dimension models~\cite{Appelquist:2000nn,Cheng:2002iz,Servant:2002hb}, vector bosons in little higgs 
models~\cite{Hubisz:2004ft,Birkedal:2006fz} and scalars in extensions of the 
SM~\cite{McDonald:1993ex,Barbieri:2006dq,Lisanti:2007ec}. Right-handed sneutrinos as a CDM candidate have also been revived lately~\cite{ArkaniHamed:2000bq,Lee:2007mt}.
Predictions for signals in direct~\cite{Servant:2002hb,Majumdar:2002mw,Cerdeno:2008ep,Barger:2007nv,Cerdeno:2004xw,Bottino:2003cz}, 
indirect~\cite{Barrau:2005au,Hooper:2002gs,Bertone:2002ms,Bergstrom:2004cy,Cheng:2002ej,Arina:2007tm,Hooper:2005fj,Mambrini:2004kv,Bertin:2002ky,
Baltz:2001ir,Gustafsson:2007pc,Perelstein:2006bq,LopezHonorez:2006gr} or collider experiments have been made 
within each of these models~\cite{Arrenberg:2008wy,Belanger:2008yc,Hooper:2007qk,Balazs:2007pf,Allanach:2007qk,
Cheng:2002ab,Baer:2005bu,Ellis:2005mb,Arnowitt:2001be,Belanger:2004ag,Djouadi:2001yk,Gomez:1999dk,
Profumo:2006bx,Matsumoto:2008fq}.
Furthermore in specific case studies, in particular  within supersymmetric models, 
the prospects of determining the properties of the new  particles and from there 
infer a "collider" prediction  for the relic density or for the detection rates were
analyzed~\cite{Allanach:2004xn,Baltz:2006fm,Belanger:2008yc,Moroi:2005nc}. 
While colliders and in particular the LHC have a good potential for  discovering and identifying
new particles present in various extensions of the standard model, direct detection experiments (DD) will be the 
ones
to provide evidence for a stable relic particle~\cite{tdr:1999fr,Ball:2007zza}.
Furthermore in some cases, direct detection experiments have better discovery prospects than 
the LHC. The best
known example is the so-called focus point region in constrained supersymmetric models~\cite{Baer:2004qq,Baer:2005ky,Feng:2000gh}.
We therefore concentrate here on direct detection
aspects and consider only models which offer the best detection prospects, those 
with a weakly interacting particle at the electroweak scale.

A number of experiments are currently searching for CDM by measuring the  elastic scattering rate on nuclei in 
large detectors. Their sensitivity is being improved and upper limits are updated regularly. 
The best upper limit  on the proton-$\chi$ spin-independent (SI) cross section has been recently obtained by 
 Xenon, $\sip<4.5 \times 10^{-8}$pb for a CDM of $30$GeV~\cite{Angle:2007uj}  and CDMS,
$\sip<4.6 \times 10^{-8}$pb for a CDM of  $60$GeV~\cite{Ahmed:2008eu}.
These limits are already  putting constraints on the parameter space of new physics models. 
Limits on spin-dependent (SD) cross-sections are much less restrictive. The best limits are now obtained by  KIMS 
for protons, $\sdp<1.6\times  10^{-1}$pb~\cite{Lee.:2007qn}  and by
Xenon 
for neutrons, $\sigma^{SD}_{\chi n}< 
6\times 10^{-3}$pb~\cite{Angle:2008we}. Indirect detection of neutrinos coming from CDM annihilation in the Sun 
sets a limit 
on $\sdp$, the best limit is  from Super-Kamiokande, $\sdp<4.\times  10^{-3}$pb~\cite{Desai:2004pq}. These  
do not yet allow to test the most popular NP models 
\footnote{ Note that DAMA/LIBRA have very recently confirmed their annual modulation signal\cite{Bernabei:2008yi}. 
We will not consider this result as it seems to be incompatible with other
searches unless the CDM particle is
below 10GeV~\cite{Petriello:2008jj}, in the sample models we consider  CDM candidates are rather in the 30-1000~GeV range.}.

One difficulty in extracting precise information from an elastic scattering rate on nuclei is that the rate
depends not only on the details of the particle physics model but also large theoretical uncertainties are 
introduced by  the CDM velocity distribution, the nuclear form factors and the coefficients that
describe the quark content in the nucleon. 
The former can be eliminated by taking ratios of rates in different materials
while a large part of the uncertainty from the quark content in the nucleon will drop out when taking ratios of proton to neutron
amplitudes. Fortunately many of the detectors set up or planned use different materials thus can be sensitive to 
different combinations of proton and neutron amplitudes. The procedure for extracting in a model independent way the amplitudes
 for spin dependent interactions on protons
and neutrons  was discussed in ~\cite{Tovey:2000mm,Giuliani:2005bd}. 
The  spin independent interactions on the other hand are basically sensitive to 
one combination of neutron and proton amplitudes,
this is because all heavy materials have roughly the same ratio of protons to neutrons~\cite{Giuliani:2005my}.
Our goal is to  see what would be the prospects for determining the  properties of CDM particles 
after a signal has been observed. We will use directly the event rates or assume that the spin (in)dependent proton and neutron amplitudes have been extracted. 
We  assume that future  spin independent and spin dependent detectors will have sufficient
sensitivity to  measure a signal. 
We will consider the maximal achievable sensitivity to be $\sip\approx 10^{-10}$~pb, for example with Eureca~\cite{Kraus:2006pj}
 and  $\sdp\approx 4\times 10^{-7}$~pb, as in the  COUPP proposal~\cite{Collar:2007xn}.
Although this last value requires a significant improvement in  SD detectors,  
we  emphasize the importance of the 
SD interactions in determining the properties of the CDM candidate.

Comparative studies  of the prospects for direct detection in new physics models have been  performed
in ~\cite{Oikonomou:2006mh,Hooper:2006xe}. Recently a comparison of the SI detection rates 
and  rates for indirect detection of neutrinos  in the case of the MUED, little Higgs and MSSM models was
 presented~\cite{Barger:2008qd}.
The potential of a combined measurements of SI and SD rates to distinguish MUED from MSSM with COUPP using
two different materials was also examined~\cite{Bertone:2007xj}. 
We expand on these analysis in many ways. First we examine a larger class of MSSM models, 
second we rely heavily on detectors sensitive to SD interactions and third we insist on the 
importance of using different materials to extract both the neutron and proton amplitudes. We also take
into account uncertainties from the quark coefficients in nucleons and use an improved calculation of
the direct detection rate~\cite{Belanger:2008sj}.
This is a first step towards a more general analysis where one would combine information 
from both direct and indirect detection as well as from collider searches, see also
~\cite{Hooper:2006xe}.

This paper is organized as follows: after setting up our notation in section 2,  we  summarize in section 3  the  predictions for the SI
 and SD cross sections on nucleons  as well as for the ratios of SD and SI amplitudes on protons and neutrons 
in different CDM models.  We consider Majorana
fermions (in particular the neutralino in SUSY),  
a right handed Dirac neutrino,  as well as vector and scalar particles.
The results of our scans over  the parameter space for each  sample models are presented in Section 4
where we show which models can in principle be distinguished by measurements
of both SI and SD amplitudes on protons and neutrons. 
The predictions for the scattering rates on various nuclei are then compared in section 5.
Finally in section 6 we briefly mention the case where a signal can be observed only in the SI interaction.
Our results are summarized in section 7.

\section{Direct detection}

The total scattering cross section of a DM particle, $\chi$,  off a point-like
nucleus for spin independent interactions reads
\begin{equation}
\sigma_{0}^{SI} = \frac{4\mu_\chi^2}{\pi} \left(\lambda_p Z
+\lambda_n (A-Z)\right)^2
\end{equation}
where  $\mu_{\chi}=m_\chi M_A/(m_\chi+M_A)$ is the reduced
$\chi$-nucleus mass and $M_A$ the  mass of the nucleus.
Te proton(neutron) amplitudes. 
are related  via some coefficients to the amplitudes for $\chi$-quark scattering, $\lambda_q$.
 For example
 for scalar interactions of Majorana fermions,  in the notation of ~\cite{Belanger:2008sj} 
 \begin{equation}
\lambda_{p,n}=\sum_{q=1,6} f^{p,n}_{q} \lambda_q\;\;\; ,
\label{eq:quark_coeff}
\end{equation}
where $f^{p,n}_{q}$  describes the contribution of quark $q$ to the mass of the nucleon.
The quark coefficients for scalar interactions 
have large uncertainties~\cite{Bottino:2001dj}.
To take these into account 
we vary the input parameters of \micromegas2.2~\cite{Belanger:2008sj} in the range 
\begin{equation}
\sigma_{\pi N}=55-73 {\rm MeV}\;\;\;{\rm and}\;\;\; \sigma_0=35\pm 5~{\rm  MeV}
\label{eq:si_coeff}
\end{equation}
 which in essence amounts to varying the s-quark  content in the
nucleon in the range $0.19<f_s^p<0.56$. The heavy quarks coefficients, $f^N_Q$, are related to those of the light
 quarks~\cite{Jungman:1995df}. 
 In the case of a Dirac fermion with an effective vectorial interaction, 
the coefficients that describe the quark content
in the nucleon just count the number of valence quarks and therefore have no
theoretical uncertainty~\cite{Jungman:1995df}. 

 For spin
dependent interactions, the point-like nucleus cross-section reads
\begin{equation}
\sigma_{0}^{SD} = \frac{\mu_\chi^2}{16\pi} \frac{J_A+1}{J_A}
\left(\xi_p S_p^A + \xi_n S_n^A\right)^2
\label{eq:SDpoint}
\end{equation}
where $J_A$ is the total spin of the nucleus and $S_{p,n}^A$ are
obtained from nuclear calculations. 
The SD  nucleon amplitudes, $\xi_{p,n}$ \footnote{Note that our definition of the nucleon amplitudes differ from the usual convention where one uses
 $a_{p,n}=\sqrt{2} G_F \xi_{p,n}$~\cite{Jungman:1995df,Tovey:2000mm}.}  are related to the quark amplitudes,  
\begin{equation}
\xi_{p,n}=\sum_{q=u,d,s} \Delta q^{p,n} \xi_q
\end{equation}
where the  coefficients  $\Delta q^{p,n}$ have  been estimated for light quarks~\cite{Jungman:1995df}  
\begin{eqnarray}
\Delta_u^p=0.842\pm 0.012; \;\;\; \Delta_d^p&=&-0.427\pm 0.013; \;\;\;
\Delta_s^p=-0.085\pm 0.018\nonumber\\
\Delta_u^n=\Delta_d^p;\;\;\;\Delta_d^n&=&\Delta_u^p;\;\;\;\Delta_s^n=\Delta_s^p
 \label{eq:axial_vector}
\end{eqnarray}
In the numerical analysis we will allow the coefficients to vary within their $1\sigma$ range.

The recoil energy distribution
measured in a detector further contains some dependence on
 the nuclear form factors as well as on the CDM velocity distribution.
\begin{eqnarray}
\frac{dN}{dE}= \frac{2M_{det} t}{\pi} \frac{\rho_0}{m_\chi}
&&\left[ F_A^2(q) \left( \lambda_p Z + \lambda_n
(A-Z)\right)^2\right.\\
&+&\left.\frac{4}{2J_A+1}\left( S_{00}(q) \xi_0^2+ S_{01}(q)
\xi_0\xi_1+S_{11}(q) \xi_1^2\right) \right] I(E)
\label{eq:rate}
\end{eqnarray}
where $\xi_1=\xi_p+\xi_n$ and $\xi_0=\xi_p-\xi_n$.  $F_A(q)$ is
the nuclear form factor for scalar interactions and $S_{ij}(q)$
the form factor for spin dependent interactions, both depend on
the momentum transfer,
$q=\sqrt{2M_A E}$~\cite{Bednyakov:2006ux}. $\rho_0$ is the local neutralino density, $M_{det}$ the detector mass 
 and I(E) the integral over the velocity distribution
\begin{equation}
I(E)=\int_{v_{min}}^\infty \frac{f(v)}{v} dv
\end{equation}
where $v_{min}=(\frac{E M_A}{2 \mu^2_\chi})^{1/2}$.
To compute the cross sections on nucleons and the event rates we rely on \micro2.2~\cite{Belanger:2008sj}.

All information on the CDM model is
contained in the amplitudes $\lambda_p,\lambda_n,\xi_p,\xi_n$
as well as in the mass of the CDM.
 Once a signal has been observed, one could use data from different detector materials 
to extract information on these amplitudes~\cite{Bourjaily:2005ax}.  
 This evidently necessitates making some
assumption about both  the halo velocity distribution, the dark
matter distribution as well as on the nucleon and nuclear form factors.
 The dependence on the velocity and on the dark matter
distribution however drops out when taking ratios of the nucleon amplitudes.
Furthermore, some of the uncertainty from the quark coefficients in the nucleon also drop out. 
This is because the sea quarks coefficients  which give the dominant contribution are identical for protons and neutrons.

We choose as independent parameters the ratios $\phi,\xi/\lambda_+,\lambda_p/\lambda_n$  and $\lambda_+$which characterizes the overall SI rate,
\begin{equation}
\xi_p=\xi \sin\phi; \;\;\; \xi_n=\xi \cos\phi; \;\;\; \lambda_+=\lambda_p+1.4\lambda_n .
\end{equation}
Note that the factor of $1.4$  in $\lambda_+$ depends on the ratio of protons to neutrons in the nucleus. Typically this ratio does not vary much~\cite{Giuliani:2005my}, our choice  gives the maximal sensitivity in heavy nuclei. 
Nuclei that are sensitive to SD interactions do so primarily through an unpaired nucleon, this means they  have either $S_p$ or $S_n\neq 0$ and 
have little sensitivity to the interference term $\xi_p\xi_n$ in eq.~\ref{eq:SDpoint}~\cite{Giuliani:2004uk}. 
The sign of $\phi=atan(\xi_p/\xi_n)$ is therefore hard to determine.

The mass of the CDM candidate can be determined  from the nuclei recoil energies. This  works best when 
$M_\chi\approx 100$~GeV~\cite{Green:2007rb} although a  new method to improve the mass determination for a heavy DM
particle using signals from two 
different detectors was proposed recently~\cite{Drees:2008bv}.

\section{DM models}

We consider a selection of models representative of different CDM
candidates in the 30GeV-1TeV range:  Majorana fermion (the neutralino in the MSSM), Dirac fermion (a right-handed neutrino), 
gauge boson (the heavy photon in little Higgs models or the $B_1$ in MUED models) or scalar particles in extended Higgs models. 
The predictions for the rates for direct detection  have been studied
in all these models and rates can vary by orders of magnitude 
within each model~\cite{Hooper:2007qk, Bertone:2004pz}. For each CDM
candidate the dominant process for elastic scattering influences
the overall scattering cross-section as well as the relations
among the proton/neutron amplitudes. We will explore these relations within
models representative of each type of CDM.
A summary of the different mechanisms for CDM
elastic scattering in various models is provided in Table~\ref{tab:direct}. 
Note that two special subclasses of the MSSM have been introduced.
The main difference between these two classes is the range of mass of squarks,
 in MSSMH they are heavy and therefore do not contribute to DD.

\begin{table}[htbp]
\caption{Dominant mechanism for CDM-nucleon elastic scattering}
\label{tab:direct}
 \begin{center}
 \footnotesize
 \begin{tabular} {|c|c|c|c|c|}
 \hline
 Model & CDM & Nature &SI & SD  \\
 \hline
 MSSMH  & Neutralino &Majorana fermion & Higgs& Z\\
MSSMQ   & Neutralino & Majorana fermion & Higgs+squark & Z +
 squark\\
 RHN & $\nu_R$ &Dirac fermion & Z +Higgs & Z \\
 MUED   & $B_1$ & Vector boson & Higgs +KK-quarks & KK-quarks \\
 LHM & $A_H$ &  Vector boson & Higgs + Quarks & Quarks \\
 IDM & $H^0$ & Scalar & Z+Higgs &  \\
 \hline
 \end{tabular}
 \end{center}
\end{table}

\subsection{ MSSM}

In the MSSM the CDM is a Majorana fermion, the neutralino, 
$\chi_1^0$
~\footnote{We do not consider the case of the sneutrino CDM which usually gives too large detection
rate unless its coupling to the Z is suppressed.~\cite{ArkaniHamed:2000bq,Lee:2007mt,Falk:1994es}} The
nature of the neutralino, whether it is mostly bino or contains a mixture of Higgsino or wino,
 strongly influences the
annihilation mechanisms and the CDM relic density. For direct
detection one gets two types of contributions, Higgs and squark
exchange for SI interactions and Z and squark exchange for SD
interactions. In general the Higgs and Z exchanges dominate since
the squark contributions suffer from a mass suppression (the squarks are generally at the TeV scale). We will consider two categories of MSSM
models. In the first, MSSMH, sfermions are  heavy (2TeV) and do not contribute to DD. In this model CDM annihilation requires a lightest neutralino with
 some Higgsino or wino component. In the second,  MSSMQ,
we force one of the squark masses to be $M_{{\tilde q}_{L,R}}<2M_\chi$. 
We introduce these classes of models as a way to quantify
the impact of the squarks in DD. Note that these two
types  of MSSM models can be easily distinguished at LHC which can probe
the squark sector up to more than 2TeV~\cite{atlas:1999fq,Ball:2007zza}.  In both cases we will assume  the sleptons  to be heavy since sleptons do  not contribute to direct detection.
However one should keep in mind that sleptons can contribute to
the CDM relic density, both through annihilation or coannihilation, 
so our analysis is not completely general when 
confined to models that are in agreement with  the measured value for the CDM relic density.
Such light sleptons can be searched for at the LHC.  

Taking into account the dominant Higgs exchange diagram only, the spin independent interaction  reads
\begin{eqnarray}
\lambda_{N}=-m_N\frac{g^2}{4 M_W c_W} \sum_{i=1,2}\left[
(f_{u}^N+ f_{c}^N+f_{t}^N)g_{h_iuu} + (f_{d}^N+ f_{s}^N+ f_{b}^N)
g_{h_idd}\right] g_{h_i\chi\chi}\frac{1}{m^2_{h_i}}
\label{eq:si}
\end{eqnarray}

\noindent where $h_i=h,H$, $g_{huu}=\ca/\sb$, $g_{hdd}=-\sa/\cb$,
$g_{Huu}=\sa/\sb$, $g_{Hdd}=\ca/\cb$ and $\alpha$ is the Higgs mixing angle. In the decoupling limit, at
large $M_H$,
$\sin\alpha=-\cos\beta$. The SUSY-QCD corrections can
shift the Higgs couplings to down-type quarks, especially at large
values of $\tan\beta$. These corrections are taken into account in
the numerical analysis bur for simplicity will be omitted from the
discussion here. The couplings of the light Higgs to $\chi_1^0$ reads
\begin{eqnarray}
g_{h\chi\chi}&=&(\ca Z_{14}+\sa Z_{13})(\cw Z_{12}-\sw Z_{11})
\end{eqnarray}
where  $Z_{1j}$ describe the field content of the LSP~\cite{Skands:2003cj}. 
Clearly for the Higgs exchange to contribute requires a LSP with some Higgsino component ($Z_{13},Z_{14}\neq0$).
In all models where  Higgs exchange dominates we expect $\lambda_p=\lambda_n$ within a 2\% accuracy. This is because the 
the quark coefficients  in protons and neutrons are the same for heavy quarks and the
largest coefficient is the one for s-quark. 

 For the spin dependent amplitude, the Z exchange contribution reads,
\begin{eqnarray}
\xi_{N}=-\frac{1}{2} (\Delta_u^N-\Delta_d^N-\Delta_s^N) \frac{g^2}{4
M_Z^2 c_W^2} (Z_{13}^2 -Z_{14}^2)
\label{eq:sd}
\end{eqnarray}
When the squark contribution is negligible the ratio of proton to neutron amplitudes $\xi_p/\xi_n$ is therefore totally independent of the neutralino 
coupling to the Z. We expect $\tan\phi=\xi_p/\xi_n\approx -1.14\pm 0.03$ when considering the range
 for the quark coefficients specified in eq.~\ref{eq:axial_vector}. 
This value for $\tan\phi$ is expected in any
model where Z exchange dominates the spin dependent interaction.

These simple relations are  
spoiled in models where squarks are
light unless squarks of different flavours are nearly degenerate  in which case
we still expect $\lambda_p/\lambda_n\approx 1$.  Strong corrections to this ratio can be found when
$m_{{\tilde q}_{L,R}} \approx m_\chi$ since in this case  twist-2 operators give a large contribution that can even cancel the leading
squark contribution~\cite{Belanger:2008sj}. Note that a twist-2 contribution, being 
proportional to the quark hypercharge, is larger for u-squarks so will
contribute mainly to the proton amplitude.
In principle the impact of light squarks is more important for spin dependent interactions
because of a possible cancellation between u and d-type quark
coefficients. However   the squark exchange is dominant for  SD cross sections that 
are usually too small to be measured even in ton-scale detectors.  
 
The ratio of SD to SI amplitudes also characterizes the model.  
When sfermions are heavy, the relic density of dark matter favours a LSP with
some wino or higgsino component. In such models
 the ratio of spin dependent to spin independent interactions
 depends strongly on the higgsino component of the LSP, eq.~\ref{eq:si} and eq.~\ref{eq:sd}, and predictions can vary in a
wide range.  

\subsection{Right-handed neutrino model (RHNM)}

A model with warped extra dimensions where the CDM is a
right-handed Dirac neutrino was proposed by
 Agashe and Servant~\cite{Agashe:2004bm}. This model contains both new fermions and new gauge bosons
 at the (multi-)TeV scale which interact mainly
 with third generation fermions. This model can be used as a
 prototype of a more general class of models with  a right-handed Dirac 
 neutrino as CDM ~\cite{Belanger:2007dx}. Whether or not there are
 additional quarks or gauge bosons, because of the large mass scale
 involved, the most important contribution to elastic scattering
 of the right-handed neutrino on nucleons is due to Z and Higgs exchange~\cite{Agashe:2004bm}.
 What is peculiar in this class of models
 is that the CDM is not a Majorana particle,
so there is an important contribution of Z exchange to both SI and SD
 nucleon scattering~\cite{Belanger:2008sj}.  Typically in this
model the elastic scattering cross sections are large and direct detection poses one
of the strongest phenomenological constraint on the model~\cite{Belanger:2007dx}.

For the dominant Z exchange contribution to the spin independent interactions 
\begin{eqnarray}
\lambda_{p}=\frac{g_Z^{\nu_R} e (1-4s_W^2)} {8M_Z^2 s_W c_W} \;\;\; {\rm and} \;\;\;
\lambda_{n}=\frac{g_Z^{\nu_R} e} {8M_Z^2 s_W c_W}
\end{eqnarray}
where $g_Z^{\nu_R}$ is the parameter that describes the coupling of $\nu_R$ to the Z. 
This coupling is induced through mixing so is suppressed with respect to the SM couplings. 
The neutron and proton amplitudes are directly related~\cite{Agashe:2004bm}, 
\begin{equation}
\frac{\lambda_p}{\lambda_n}= (1-4s_W^2)  \approx 0.09 
\label{eq:nur_si}
\end{equation}  for $s_W^2 =0.228$. 
If the $\nu_R$ also couples to  the Higgs, both $\lambda_p$ and $\lambda_n$ will receive the same
additional contribution, thus modifying the simple relation, eq.~\ref{eq:nur_si}.  
In the numerical analysis we will
include a generic coupling of the Higgs to the neutrino $g_H$~\cite{Belanger:2007dx}. 

For spin dependent interactions which also proceed through Z exchange, we get
\begin{equation}
\xi_{N}=\sum_{q=u,d,s} b'_q \Delta_q^{N} \;\;\; {\rm where} \;\;\; b'_q=\frac{g_Z^{\nu_R}
(g_R^q-g_L^q)} {4M_Z^2}
\end{equation}
and $g_R^d-g_L^d= -(g_R^u-g_L^u)= \frac{e}{2s_W c_W}$. As for the MSSM, the ratio of proton to neutron amplitudes, 
$\tan\phi=\xi_p/\xi_n=-1.14\pm 0.03$.  

In the limit that the Higgs contribution is negligible, the ratio of SD  
to SI amplitudes is also independent of the details of the model 
 with  $\xi/\lambda_+=1.06\pm 0.02$ when varying 
the quark coefficients in the range specified in eq.~\ref{eq:axial_vector}.

\subsection{Universal extra-dimensions model (MUED)}

In the universal extra dimension model (UED) potential dark matter candidates include a KK gauge boson, a
KK neutrino, a KK scalar  or a KK graviton~\cite{Dobrescu:2007ec,Servant:2002aq,Cheng:2002ej}.
We restrict our analysis to  the minimal UED model (MUED), in which case 
the CDM 
is either the first KK level of the hypercharge gauge boson, $B^1$ or the KK graviton.
We will consider only the former possibility
 since the graviton  has small direct detection rates.  
CDM scattering on nucleon proceeds both through Higgs exchange and KK-quark exchange. 
For spin independent interactions, the nucleon amplitude reads~\cite{Servant:2002hb}
\begin{equation}
\lambda_{N}=\frac{m_N}{8 m_{B_1}} \sum_q\left( \frac{g_1^2}{m_h^2}+ 2g_1^2(Y^2_{q_L}+Y^2_{q_R})
\frac{M^2_{B_1}+M^2_{q^1}}{(M^2_{B^1}-M^2_{q^1})^2}\right)  f_q^N
\label{eq:si_ued}
\end{equation}
where the sum is over all quark flavours and $g_1=e/c_W$, $Y_{q_L}=1/6,Y_{u_R}=4/6,Y_{u_R}=-2/6$. The first term arises
from Higgs interactions and the second term from KK quarks exchange. 
In this model it is quite natural to have a large contribution 
from KK-quarks since they are  nearly degenerate with the CDM. 
We include radiative corrections to level 1 KK states~\cite{Cheng:2002iz}  
which lead to  mass splittings between KK quarks and $B^1$.
Note that the Higgs contribution is suppressed  compared
 to the MSSM by a factor
$m_W/m_{B_1}$ as well as by the  Higgs mass which is usually larger than in the MSSM.  
Nevertheless one expects  $\lambda_p/\lambda_n \approx 1$ as in models where the Higgs exchange dominates
because all new quarks are nearly degenerate. 

For spin dependent interactions, the amplitude reads
\begin{eqnarray}
\xi_N&=&  \frac{1}{\sqrt{6}}\sum_{q=u,d,s} 2g_1^2(Y^2_{q_L}+Y^2_{q_R})
\frac{1}{(M^2_{q^1}-M^2_{B^1})} \Delta_q^N 
\label{eq:sd_ued}
\end{eqnarray}
and is solely due to KK quarks exchange. 
One can easily show that 
 $\tan\phi=\xi_p/\xi_n\approx -3.5 $ independently of the parameters of the model as long as all KK quarks are degenerate. 
 The ratio $\xi/\lambda_+$ can be  large and is 
 controlled by the $B_1$ mass and by the mass splitting with the KK-quarks when these dominate the SI interaction.

\subsection{Little Higgs model (LHM)}

In the little Higgs model with T-parity,  the dark matter candidate is the lightest new heavy neutral
gauge boson $A_H$ ~\cite{Cheng:2003ju,Cheng:2004yc}. This model therefore shares many aspects of the MUED model just discussed,
the CDM is a gauge boson, spin independent interactions are due to Higgs and heavy quark exchange, 
while
only the latter contributes to spin dependent interactions. The expressions for both SI and SD amplitudes
are the same as above, eqs.~\ref{eq:si_ued},\ref{eq:sd_ued}.
There are however two important differences between these models: 
first the  hypercharges  of the  heavy quarks are small, $Y_{q_L}=1/10,Y_{q_R}=0$. Second 
 the mass splitting between the new heavy quarks $Q$
 and $A_H$ is typically
much larger than in the MUED model, which means that the heavy quark contribution to DD  is suppressed.
One therefore expects an overall low rate $\ssi$  and $\lambda_p/\lambda_n\approx 1$ 
when Higgs exchange dominates or heavy quarks are nearly degenerate. Because there is no Z exchange diagram the
SD interaction should also be much suppressed unless one artificially requires a small mass splitting between
the heavy photon and heavy quarks, eq.~\ref{eq:sd_ued}. It is only in this case that one expects 
to have a detectable cross section. 
Then $|\phi|$ will depend strongly on the mass difference between the heavy photon and the lightest new quark
and should be large.

\subsection{Scalar dark matter }

Simple models with an additional scalar field that is basically decoupled from the SM sector have
been proposed~\cite{McDonald:1993ex,Barbieri:2006dq,Lisanti:2007ec}. In these models, the CDM candidate is  a new scalar field. 
We consider the Inert Doublet Model(IDM)~\cite{Barbieri:2006dq}, a two Higgs doublets extension of the standard model with  a $Z_2$ symmetry. 
One of the two doublet and the usual standard model particles are even under this symmetry. The new particles of the model are a neutral
 ($H^0$), a pseudo ($A$) and a charged ($H^+$) scalar. Depending on the parameters of the model, the  dark matter candidate can be $H^0$ or $A$.
Only spin independent interactions can occur  through either   $H^0q\xrightarrow{h}H^0q$ and 
$H^0q\xrightarrow{Z}Aq$. 
The latter has to be kinematically forbidden,
 that is  $M_A-M_{H^0} > 100$~keV,  to respect the current experimental constraints.
The $h$ exchange cross section
driven  by an effective coupling $\lambda_L$ ~\cite{Barbieri:2006dq} is :
\begin{equation}
\sigma_{\chi N}=\frac{\mu_\chi^2}{4 \pi}\left(\frac{\lambda_L}{M_{H_0}M_h^2}\right)^2 (\sum_q f^N_q)^2 m_N^2,
\label{eq:sigmaDMpIDM}
\end{equation}
where $\chi=H^0$.

\section{Results}

Here we present numerical results for each  of our sample models. Amplitudes and cross sections for direct detection are computed with
\micromegas2.2 and in each case include all tree-level diagrams, the contribution of twist-2 operators as well as QCD
corrections. Additional SUSY-QCD corrections are included in the MSSM as discussed in ~\cite{Belanger:2008sj}.
The computation of the CDM relic density is also based on ~\micromegas2.2~\cite{Belanger:2006is,Belanger:2004yn}. We fix $m_t=172.6$~GeV. 
We also restrict the parameter space to a CDM particle roughly below the TeV scale simply because detectors are not as sensitive to  heavier
CDM particles. We
also never consider $m_h>500$~GeV, although allowed in some models such a Higgs gives a small contribution to  direct detection.

We first summarize for each model the predictions for both $\ssi$ and $\ssd$.
We always impose the upper limit from the relic density of dark
matter $\Omega h^2<0.136$~\cite{Hamann:2006pf} in our scans as well as other model dependent  constraints on the parameters of each model. 
When specified we also impose the lower bound $\Omega h^2>0.094$~\cite{Hamann:2006pf}.
We then compare the ratio 
of amplitudes on neutrons and protons ($\phi,\xi/\lambda_+,\lambda_p/\lambda_n$) before
comparing the rates on various nuclei. We have taken into account the theoretical uncertainty in the
coefficients that relate the amplitude for quarks to the one in  nucleons by varying the input parameters in 
the range specified in eqs.~\ref{eq:si_coeff},\ref{eq:axial_vector}.

\subsection{Predictions for $\ssi$ and $\ssd$}

\subsubsection{MSSM}

We consider two specific classes of the generic MSSM, as mentioned above. 
In the first class, MSSMH, squarks
are heavy and  we assume only universality among two of the gaugino masses at the GUT scale,
that is $M_3=3M_2$ at the weak scale.  
 In the second class, MSSMQ, we allow for light squarks, for simplicity we also 
impose full universality of the gaugino masses, which leads at the weak scale to  $M_3=3M_2=6M_1$ .
 In all cases we assume heavy sleptons.

For each model we have scanned over $10^5$ scenarii varying  the model parameters defined 
at the weak scale in the range 
\begin{eqnarray}
&&100~{\rm GeV}<M_1<1000~{\rm GeV}; \;\;\; 100~{\rm GeV}<\mu<2000~{\rm GeV} 
\nonumber\\ 
&&100~{\rm GeV}<m_A<2000~{\rm GeV}; \;\;\;
2<\tan\beta<52\;\;\;
\end{eqnarray}
and
\begin{eqnarray} 
MSSMH &:& \;\;\; 100~{\rm GeV}<M_2<1000~{\rm GeV} \;\; ; \;\; M_{\tilde q_{L,R}} =2~{\rm TeV}\nonumber\\ 
MSSMQ &:& \;\;\; M_2=2M_1 ; \;\; \; M_{\tilde q_{L(R)}} <2M_\chi; \;\;  M_{\tilde q_{R(L)}} =2~{\rm TeV} 
\end{eqnarray} 
 For this range of parameters the mass of the DM particle does not exceed $1$~TeV.
In each case the LEP limits on Higgs and SUSY particles are imposed as well as the upper bound on the CDM relic density.

\begin{figure}
\setlength{\unitlength}{1mm} \centerline{\epsfig{file=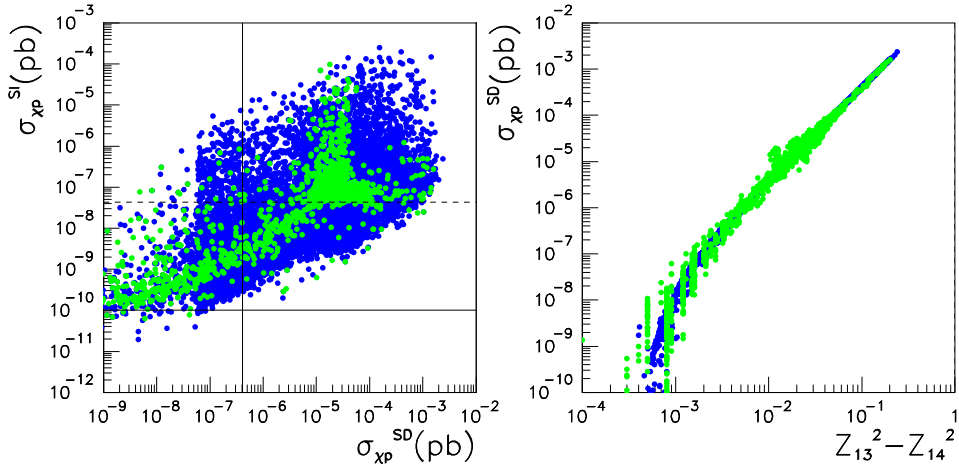,width=14cm}}
\vspace{-1cm}
  \caption{a) Predictions for $\sip$ vs $\sdp$ in MSSMH. In blue  the scenarios that satisfy the WMAP upper bound  and in green 
those that have $0.094<\Omega h^2<0.136$.  For easy reference the present absolute lower limit from CDMS/Xenon is indicated (dash) as
 well as future limits from large scale detectors (full) b) $\sdp$ as a function of $Z_{13}^2-Z_{14}^2$ in MSSMH. }
     \label{fig:MSSMH}
\end{figure}

The predictions for the SI and SD cross-sections in MSSMH are displayed in fig.~\ref{fig:MSSMH}a together with the reach of future
ton-scale detectors. The  absolute bound for CDMS/Xenon is indicated only to guide the eye as this limit depends on the CDM mass.
Virtually all scenarios will be accessible to future searches for SI interactions. This is a direct
consequence of imposing the constraint from the relic density which requires a neutralino DM  with some higgsino 
component for efficient annihilation. This then automatically leads to a $\chi_1^0$ coupling to the light Higgs 
hence to a non negligible cross section for SI elastic scattering. The smallest cross-sections in fig.~\ref{fig:MSSMH}a correspond to a neutralino with a very small higgsino fraction
that nevertheless annihilate efficiently because it does so near a heavy Higgs resonance.
Models with a relic density within the WMAP range (rather than only below the upper bound)  almost span the full range of 
predictions for SI and SD cross sections, 
although many of the  scenarios with the largest $\ssi$ have  a large higgsino component and
are  associated with a small value for the  relic density due to the efficient annihilation into W pairs.
The higgsino component also induces a coupling to the Z hence leads to SD interactions. Because these
interactions are not coherent in several cases the predictions can be as low as $\sdp\approx 10^{-9}$~pb, much
below the expected reach of future detectors. Note that because Z exchange dominates SD interactions, the rate is directly related to the $Z\chi\chi$ coupling which is proportional to
$Z_{13}^2-Z_{14}^2$, eq.~\ref{eq:sd}. A rate measurement will therefore set a limit on this
coupling assuming the MSSMH, see fig.~\ref{fig:MSSMH}b. 

\begin{figure}
\setlength{\unitlength}{1mm} \centerline{\epsfig{file=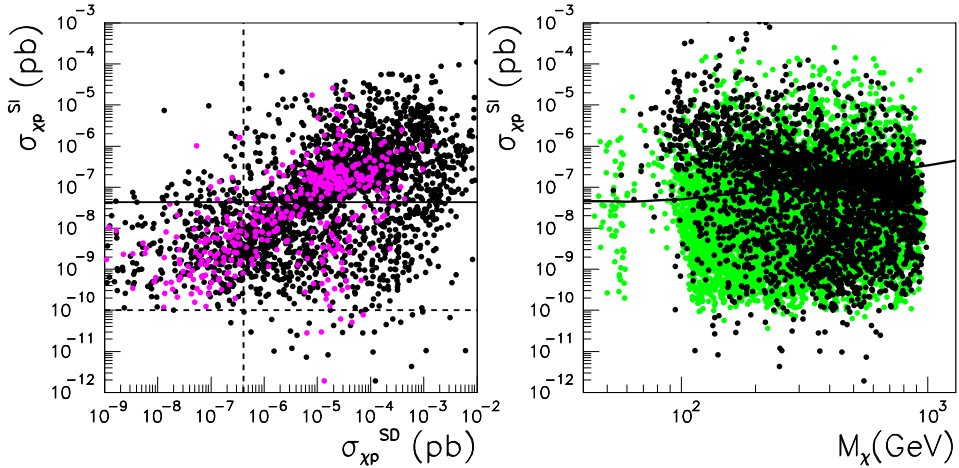,width=14cm}}
\vspace{-1cm}
  \caption{a) Predictions for $\sip$ vs $\sdp$ in MSSMQ. In black the scenarios that satisfy the WMAP upper
   bound  and in green 
those that have $0.094<\Omega h^2<0.136$.  For easy reference the present absolute lower limit from CDMS/Xenon is indicated (full) as
 well as future limits from large scale detectors (dash) b) $\sip$ as a function of the neutralino mass in  MSSMQ (black) and MSSMH(green)}
     \label{fig:MSSMQ}
\end{figure}

In MSSMQ, the range of predictions for $\sip$ is roughly the same as in MSSMH, see fig.~\ref{fig:MSSMQ}a although  cross
sections below  the reach of future SI detectors can be expected in a few cases. Furthermore
large cross sections for SD interactions can be expected even when SI ones are quite
low.  In general this occurs in  scenarios with light squarks. There is no explicit correlation between $\ssi$ and the mass 
of the CDM, see fig.~\ref{fig:MSSMH}b, 
although models with a neutralino around $60$~GeV that annihilate near a light Higgs resonance can have a  small cross section. 
We explicitly display in fig.~\ref{fig:sigma_models}b, the CDMS exclusion limit for both MSSMH and MSSMQ. 
Many scenarios are excluded even when taking into account  a large
uncertainty (up to  a factor 3~\cite{Bottino:2001dj}) in the exclusion limit that could arise from the DM   
distribution.

\subsubsection{RHNM}
In the right-handed neutrino model we use as free parameters the mass of the CDM, its coupling to the Z, $g_Z^{\nu_R}$, and to the Higgs, $g_H$, 
as well as the mass of the Higgs. We assume all other particles in the model to be above 3~TeV and therefore  
do  not play a role in direct detection.  We perform a scan over 100000 models varying the  free parameters  in the range
\begin{eqnarray}
&&30~{\rm GeV}<m_{\nu_R}<1200~{\rm GeV}\;; \;\;\;  120~{\rm GeV}<m_h<500~{\rm GeV}\nonumber \\
&& 0.001<g_Z^{\nu_R}<0.01\;; \;\;\; 0.01<g_H<0.25
\end{eqnarray}
The range of $g_Z^{\nu_R}$ is chosen so that the upper bound on $\Omega h^2$ is easily satisfied while not giving too large
$\ssi$ whereas the range for the  Higgs coupling $g_H$ is set so that the Higgs can potentially play a role in DD.
Models with a Dirac right-handed neutrino  often have an extended gauge sector. Since this  is mostly relevant for the annihilation of a CDM particle
beyond the   TeV scale,  we can safely ignore this sector in our analysis.

In this model one expects rather large rates for SI interactions, and this is in fact the most severe constraint on the model.
Furthermore a strong correlation  is expected between  the SI and SD rates as seen in fig.~\ref{fig:sigma_models}a. Indeed these rates
  are governed 
by the Standard Model axial and axial-vector $Z\bar{q}q$ couplings.
Note that the limit extracted from $\ssi$ has to be rescaled to take into account the fact that in this model 
$\lambda_p\ll\lambda_n$. In  practice  it means rescaling the
limit by a factor 2-3 depending on the material.  
Even taking this factor into account fig.~\ref{fig:sigma_models}a shows that models that will  not be excluded 
in the near future predict a low rate for $\sdp$.
The mass of the CDM allowed in this model is either near $M_Z/2$, $M_H/2$ 
or above 500~GeV~\cite{Belanger:2007dx}.

\subsubsection{MUED}

For the computation of the direct detection rate in the MUED model we include the  
level one  KK quarks as well as the lightest Higgs exchange. We ignore 
the level  2 Higgs since in elastic scattering cross sections a heavy Higgs suffers from a mass suppression. Furthermore the coupling of the level 2 Higgs to the $B^1$ is    loop induced hence suppressed.
Note however that because $M_{H^2}\approx 2M_{\chi}$ the level 2 Higgs  plays a role in the computation of the relic density~\cite{Kakizaki:2006dz,Kakizaki:2005en}. The impact of neglecting this coupling 
on our analysis is not significant as the majority of the models already 
has $\Omega h^2<0.136$.

The free parameters of the model are 
$1/R$, the inverse size of the extra dimension that determines the mass of the  KK states,
$\Lambda$, the cutoff scale and $m_h$ the lightest Higgs mass. 
We scan over $10^5$ scenarii 
with the three free parameters of the model in the following range
\begin{equation}
300~{\rm GeV}<1/R<1300~{\rm GeV}; \;\;\; 3<\Lambda R<50; \;\;\; 120~{\rm GeV}<m_h<500~{\rm GeV}
\end{equation}
The precision electroweak constraints set
the lower bound on $1/R$~\cite{Appelquist:2002wb} while 
perturbativity and unitarity constraints set a range for $\Lambda R$~\cite{Chivukula:2003kq}.
The mass of all KK states
are computed including one-loop corrections~\cite{Kong}. The radiative corrections  will induce a small mass splitting 
between the level one $B^1$ boson and KK-fermions. Such 
splitting is typically
 1-2\% for KK-leptons and  $~5-25\%$ for KK-quarks and   
strongly influences the direct 
detection rate. We also insure that  the CDM relic density  satisfies the WMAP upper bound and that the 
charged Higgs is not the CDM.

\begin{figure}
\setlength{\unitlength}{1mm} \centerline{\epsfig{file=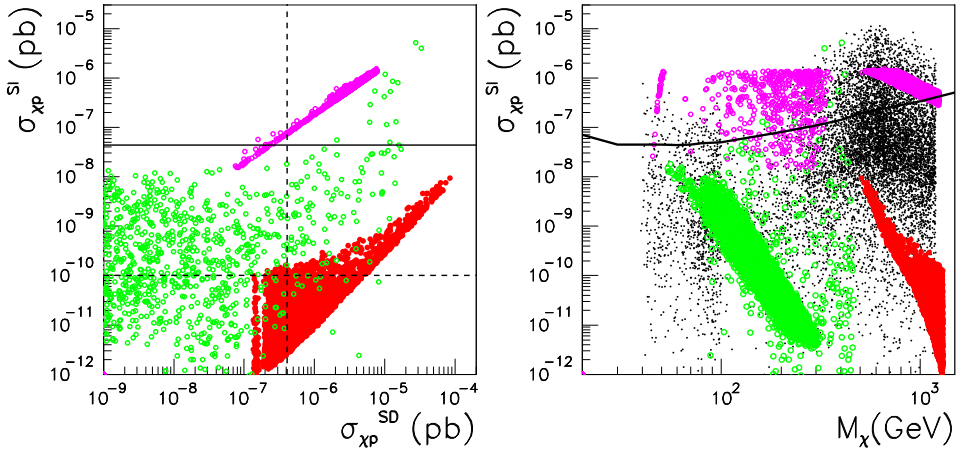,width=14cm}}
\vspace{-1cm}
  \caption{Predictions for $\sip$ vs $\sdp$ in MSSMQ(black), MUED (red), LHM (green), RHNM (pink). 
b) $\sip$ as a function of the CDM mass, same color code as a) with in addition the model IDM
 (black) }
\label{fig:sigma_models}
\end{figure}

The SI cross sections are suppressed by the heavy $B_1$ mass, eq.~\ref{eq:si_ued},
the larger cross sections are therefore expected for
the lighter CDM particles, see fig.~\ref{fig:sigma_models}b. Typically, more than an order of magnitude improvement in  detectors sensitivities
is needed  to probe the parameter space of the model and  a large fraction of the models, specially those with a CDM at the TeV scale,  
will remain inaccessible to the large scale detectors.
The main characteristic of this model is the correlation between SI and SD cross sections, this is because the heavy KK-quark exchange 
contributes to both modes.
As a result SD interactions could be  accessible in cases where rates are too low for SI interactions. This is in sharp contrast with the MSSMH. 

\subsubsection{LHM}
The LHM with T-parity contains in addition to heavy gauge bosons,  heavy T-odd fermions as well as a new T-even heavy 
top quark. 
We choose as  free parameters the Higgs mass,  $f$, $ \kappa$ and $s_\alpha$.
$f$  sets the scale of the heavy gauge bosons and
fermions in particular  the  heavy photon of mass
\begin{equation}
M_{A_H}=\frac{g' f}{\sqrt{5}}\left[  1-\frac{5v^2}{8f^2}\right]  
\end{equation}
with $v$ the usual vev of the Higgs.  $\kappa$ is an additional parameter that enters the fermion masses, for example for a heavy down-type quark,
 $M_d=\sqrt{2} \kappa f$. For simplicity we assume an 
universal factor $\kappa$ for all heavy fermions. $s_\alpha$ depends on the
ratio of the Yukawa couplings of T-even and T-odd top quarks.~\cite{Belyaev:2006jh}
 This parameter enters the top quark mass 
as well as couplings involving standard and heavy top quarks. 

We scan over $10^5$  scenarii varying randomly the
free parameters in the range
\begin{eqnarray}
&&500~{\rm GeV}<f<3000~{\rm GeV} ;\;\;\; 120~{\rm GeV}<m_h<500~{\rm GeV}\nonumber\\
&& 0.1<s_\alpha<0.96; \;\;\; 0.11<\kappa<1
\end{eqnarray}
 We impose the LEP limits on  the production of heavy quarks
 as well as on the Higgs mass.  

The rates for both SI and SD cross sections are in general quite low, even below the scale in 
fig.~\ref{fig:sigma_models}a. As we have explained before this is
due to the small hypercharge of the heavy quarks
 as well as to their generally  large mass.
Models that could lead to a signal in either the  SI or SD  channel are those where the
mass splitting between  heavy quarks and the heavy photon is between
1-10\% or slightly larger if the heavy quarks are around 100~GeV. Furthermore the heavy photon has to be rather light with  
$M_{A_H}<400$~GeV, see fig.~\ref{fig:sigma_models}b.
A Higgs near the lower LEP limit also helps
increasing the signal for SI interactions. Note that since it is
 the new quarks that couple to the nucleon 
that need to be light, mainly the first and second generation, 
the recent Tevatron limit on the heavy top quark~\cite{Carena:2006jx} does not play a role here.

\subsubsection{IDM}
In the IDM, the free parameters are those of the Higgs potential \cite{Barbieri:2006dq}. 
We choose to use rather the physical parameters, the masses of  the CDM candidate, $m_{H^0}$, the light scalar, 
$m_h$, the  pseudoscalar, $m_{A}$ and the charged higgs, $m_{H^+}$ as well as two parameters of the Higgs potential
$\mu_2$ and $\lambda_2$. Our numerical results are not very sensitive to the value of $\lambda_2$ so for
simplicity we fix $\lambda_2=0.1$. Other free parameters are varied in the range   
\begin{eqnarray}
10~{\rm GeV}< m_{H^0}<1200~{\rm GeV};\;\; 115~{\rm GeV}<m_h<500~{\rm GeV};\;\; 10~{\rm GeV}<\mu_2<1200~{\rm GeV}\;\;\;
\end{eqnarray}
with in addition the following range for the mass differences 
\begin{eqnarray}
5{\rm GeV}<m_A-m_{H^0}<15{\rm GeV};\;\;\; 40~{\rm GeV}<m_{H^+}-m_{H^0}<50~{\rm GeV}
\end{eqnarray}
if $m_{H^0},\mu_2<100$~GeV, otherwise
\begin{eqnarray}
 3{\rm GeV}<m_A-m_{H^0}<6{\rm GeV};\;\;\; 5~{\rm GeV}<m_{H^+}-m_{H^0}<10~{\rm GeV}.
\end{eqnarray}
We impose the following constraints on the model : vacuum stability and perturbativity conditions on the potential parameters,
 LEP limit on the charged Higgs, contribution to the Z boson width and electroweak precision constraints ~\cite{LopezHonorez:2006gr}.
 
 The rates for $\ssi$ varies over several orders of magnitude and  the  masses of the CDM particle ranges anywhere from
 $50$~GeV to the TeV scale, fig.~\ref{fig:sigma_models}b. Note however
 that once one imposes a lower bound on $\Omega h^2$, the ranges for the  masses
 and the direct rates are severely restricted, see section~\ref{sec:sionly}.

\subsection{Discriminating models : amplitudes for scattering on protons and neutrons}

The ratios of proton to neutron amplitudes  apart from being
free of large theoretical uncertainties provide a good model discriminator. For this to be useful
one has to assume that these quantities can be measured, this means that in this
section we keep only models for which
$\ssi>10^{-10}$~pb and $\ssd>4.\times 10^{-7}$~pb.  The more challenging case with
 only a detectable SI cross section will be discussed in the subsection~\ref{sec:sionly}.

\begin{figure}[htb]
\vspace{-1cm}
\setlength{\unitlength}{1mm} \centerline{\epsfig{file=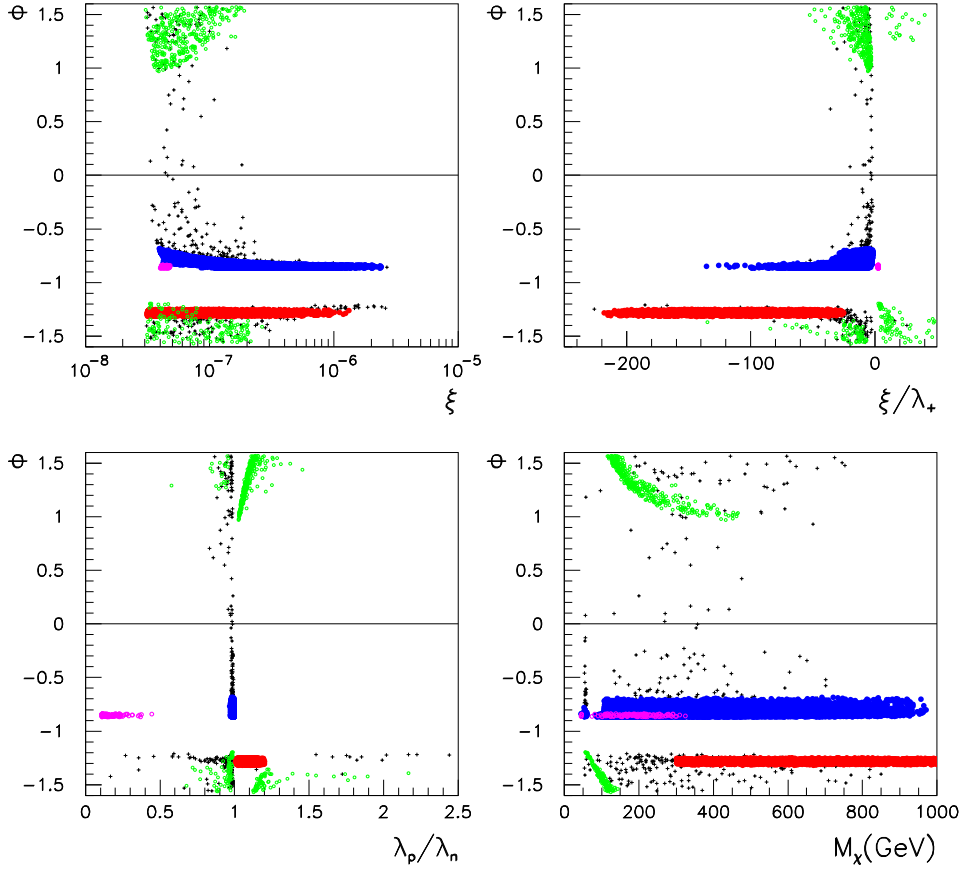,width=14cm}}
\vspace{-1cm}
  \caption{Predictions for a) $\phi=atan(\xi_p/\xi_n)$ as a function of
 $\xi$ b) $\xi/\lambda$ c)$\lambda_p/\lambda_n$  d) $M_\chi$ 
in models MSSMH(blue), MSSMQ(black), MUED(red), LHM(green) and RHNM (pink). 
Only models for which $\sip>10.^{-10}$~pb and $\sdp$ or $\sdn >4.^{-7}$~pb are included.}
 \label{fig:amp}
\end{figure}

The results of the parameter scan  for  the five models under consideration 
are displayed in fig.~\ref{fig:amp}a-c for $\phi=\arctan(\xi_p/\xi_n)$ vs $\xi$, $\xi/\lambda_+$ and  
$\lambda_p/\lambda_n$.
The ratio of SD neutron to proton amplitudes, $\tan\phi$, can discriminate models where the SD interaction is dominated by Z exchange (MSSMH and RHNM) from those where it is dominated by (s)quark exchange (LHM, MSSMQ and MUED).
The ratio of SD/SI amplitudes, $\xi/\lambda_+$ which can be much larger in the MUED or in the MSSM 
could provide further discrimination. 
The parameter $\lambda_p/\lambda_n$  can in principle disentangle
further some  models where the SI interaction is dominated by H exchange or by  Z exchange (RHNM), see fig.~\ref{fig:amp}c. Unfortunately in practice different materials are not very sensitive to this quantity. 
Note also  that the effect of twist-2 operator in MSSMQ and LHM  can be important 
and lead to large  corrections to the expected value of $\lambda_p/\lambda_n\approx 1$.

The mass determination in DD experiment from the shape of the energy spectrum  could in some cases provide additional information to discriminate between models. In particular one could  distinguish the LHM, 
which allows a  CDM in the range $M_{A_H}\approx 50-120$~GeV from MUED which requires a heavy CDM particle and even sometimes from the MSSMQ which predicts a large range for the masses fig.~\ref{fig:amp}d. 
The LHC, with its potential for discovery of coloured particles, will establish whether or not 
coloured particles  could play a role in direct detection. Indeed in all LHM predicting a signal in DD or in the MSSMQ models, the heavy quarks and squarks can be produced easily as they lie well below 2TeV. The heavy quarks can be just beyond the LEP exclusion bound, $M_{Q_H}=100-400$~GeV in the LHM while in the MSSMQ models that have a large value for
$|\phi|$  squarks can be as heavy as $M_{\tilde q}<900$~GeV. The heavier squarks occur when the neutralino has a large higgsino content. We will not pursue a detailed analysis of what can be measured at LHC, this is
beyond the scope of this paper. We note however that the mass splitting is an
issue as regards the LHC potential for discovering new coloured particles, for small mass splitting the  signals for the new particles will be hard to extract from the background. 
This could be crucial for the LHM where the mass splitting between the CDM  and the heavy quark is below 10\%.  In the MUED model
the mass splitting  is between 6-22\% and has been shown to be sufficient for providing a signal in four leptons + missing energy channel
~\cite{Arrenberg:2008wy}. In MSSMQ the mass splitting is also typically around 20\%. 
To completely discriminate between the models at the LHC  would however also require spins of the new coloured particles 
 to be measured~\cite{Datta:2005zs}.

\subsection{Direct detection rates on nuclei}
\begin{figure}[htb]
\setlength{\unitlength}{1mm} \centerline{\epsfig{file=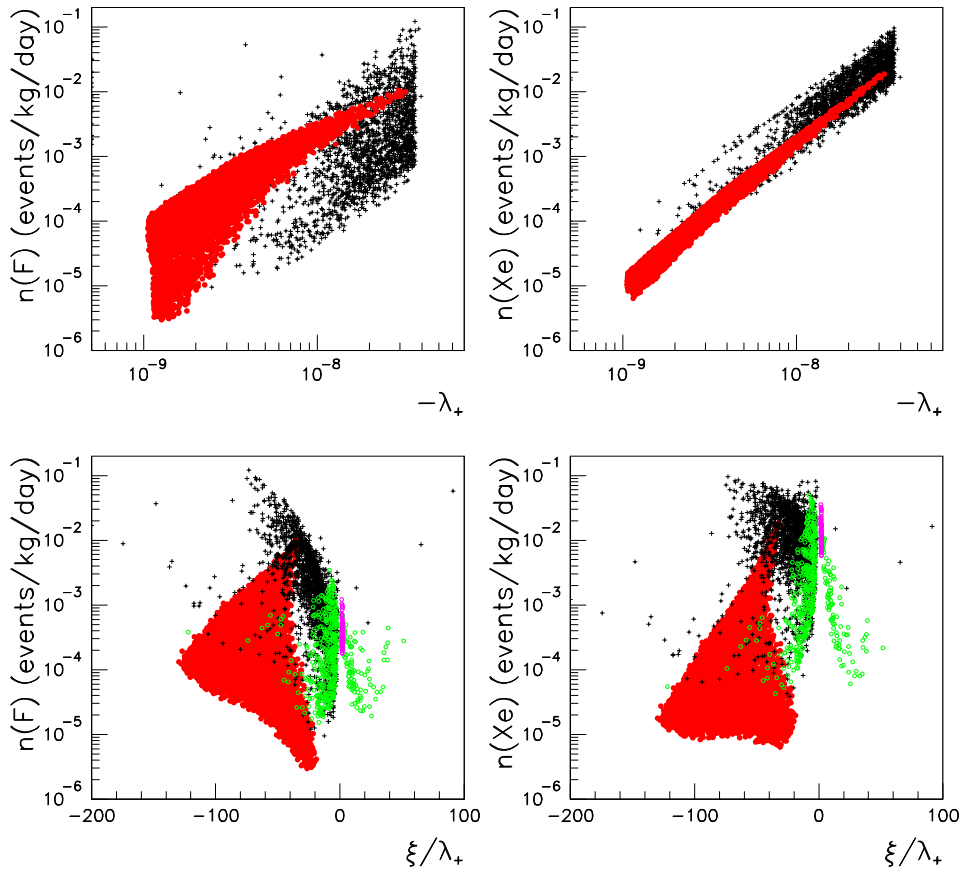,width=14cm}}
\vspace{-1cm}
  \caption{Total rates (events/kg/day) 
 a) $n(F)$ and  b)  $n(Xe)$ vs $\lambda_+$ 
c)   $n(F)$ and d) $n(Xe)$ vs $\xi/\lambda_+$   in models MSSMQ (black) and 
 MUED(red).
Predictions for rates in LHM (green) and RHNM (pink) are also included in c-d. 
Only models for which $\sigma^{SI}>1.\times 10^{-10}$~pb and $\sigma^{SD}>4.\times 10^{-7}$~pb are included. }
 \label{fig:rates}
\end{figure}

Having established that  a combined measurement of the amplitudes for  SI and SD interactions 
on protons and neutrons can in principle distinguish between the underlying particle physics models, up to a few ambiguities,
we now compare various models predictions for
quantities that are closely related to the observables.  As before we only include scenarios that could 
eventually lead to a signal in a large detector in both the SI and SD mode.  
We consider  a selection of nuclei that are currently used in large detectors. 
Those include the  nuclei sensitive only to SI interactions such as $^{40}Ar,^{76}Ge$
as well as nuclei with an odd nucleon that are also sensitive to SD interactions on either protons, 
$^{19}F,^{23}Na,^{127}I,^{133}Cs$, or neutrons
$^{29}Si, ^{73}Ge,^{129}Xe,^{131}Xe$. For each nucleus $N$ we compute the total event rate, $n(N)$ 
for a recoil energy above $2$~keV. 
For heavy nuclei, say $^{129}Xe$,  the rate is correlated with the value of 
$\lambda_+$ while for light nuclei like $^{19}F$ the correlation is spoiled by the  
SD contribution. This is illustrated in fig.~\ref{fig:rates} for both MUED and MSSMQ. 
The total number of events varies over several order of magnitude for these two
models. The rates  are generally expected to be larger for heavy nuclei, 
especially in models with a suppressed SD contribution like RHNM. In this model the total number of events in $n(F)$ varies between 
$0.4-4.\times 10^{-3}$(events/kg/day).

To eliminate as much as possible the astrophysical and 
nucleon ambiguities we compare  ratios of rates for scattering  on different nuclei. We define the ratios 
$R_{N_1/N_2}=n(N_1)/n(N_2)$. 
We  first compute the   ratios $R_{N_1/N_2}$ for  SD proton sensitive over SD neutron sensitive nuclei in our sample models. 
Such ratios are  expected to feature a dependence on  $\tan\phi=\xi_p/\xi_n$ as well as on $\xi/\lambda_+$ when a heavy nuclei is involved.

\begin{figure}
\vspace{-1cm}
\setlength{\unitlength}{1mm} \centerline{\epsfig{file=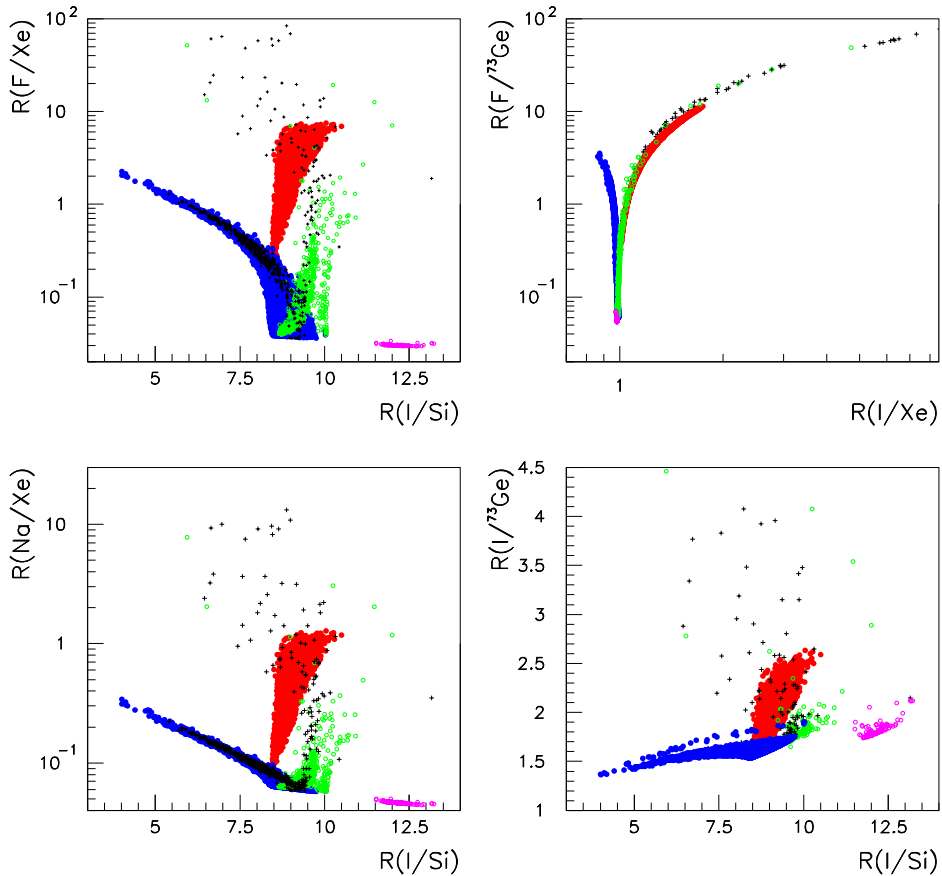,width=14cm}}
\vspace{-1cm}
  \caption{Predictions for the  ratio of total rates  
a)   $R_{F/Xe}$ vs $R_{I/Si}$, b) $R_{F/Xe}$ vs $R_{I/^{73}Ge}$, c)$R_{Na/Xe}$ vs $R_{I/Si}$, 
d)   $R_{I/^{73}Ge}$ vs $R_{I/Si}$  in different models, same color code as fig.~\ref{fig:amp}.} 
 \label{fig:nucsd}
\end{figure}

The results of a scan over the parameter space of each of our sample 
models are displayed in 
fig.~\ref{fig:nucsd}a.  
The comparison of $R_{F/Xe}$~\footnote{For illustrative purposes we use $^{129}Xe$ in the figures, similar results are found
for $^{131}Xe$.} and $R_{I/Si}$  provides a good model discriminator 
with  in particular only a small overlap between the predictions of MSSMH and MUED.
This is a direct consequence of the large value for $\tan\phi$ and $\xi$ in MUED. In this model,
$^{127}I$  is very sensitive to SD interactions since $\xi_p$ is enhanced, thus 
$R_{I/Si}$ is determined by the SD interaction and is large. On the other hand in MSSMH, 
$^{127}I$ is mostly sensitive to SI interactions thus $R_{I/Si}$ can be reduced significantly in the scenarios where $\xi$ is
large. Recall that those are the scenarios with a large  Higgsino fraction.
 Such scenarios are precisely those that lead to a  value for
$R_{F/Xe}\tilde {\cal O}(1)$  and that could have been confused with MUED. Indeed in MUED  
most scenarios predict $R_{F/Xe}>1$ because of an important SD amplitude.  
The LHM and MSSMQ scenarios that predict large values for $\tan\phi$, those with a (s)quark that is almost degenerate with the CDM particle have predictions similar to MUED for both ratios.
 Note that in the MSSMQ when the squark contribution is important there can be a partial 
 cancellation between the various quarks contributions in the neutron amplitudes 
 such that $\xi_n\ll \xi_p$ ($\phi\approx -\pi/2$). Then $n(F)$ and to a lesser extent
 $n(I)$ are enhanced but not $n(Si)$, these scenarios correspond to the few points in  fig.~\ref{fig:nucsd}a 
 with large $R(F/Xe)$ and large $R(I/Si)$. 
$^{23}Na$ is another light nuclei that shares most of the characteristics of $F$ albeit with a reduced
sensitivity to the SD part, 
see fig.~\ref{fig:nucsd}c while $^{73}Ge$ is a heavy nuclei that share many of the features of $Xe$. The ratio
 $R_{F/^{73}Ge}$ spans roughly the same range  than  $R_{F/Xe}$ in each model, see fig.~\ref{fig:nucsd}b.
The ratio $R_{I/^{73}Ge}$ just as $R_{I/Xe}$ does not  vary much  in either MSSM or LHM, except in the special scenarios with much enhanced $\xi_n$. In MUED, $R_{I/^{73}Ge}$  goes from $1.5-3.5$ with  high values associated with large $\xi/\lambda_+$, see fig.~\ref{fig:nucsd}d.

\begin{figure}
\setlength{\unitlength}{1mm} \centerline{\epsfig{file=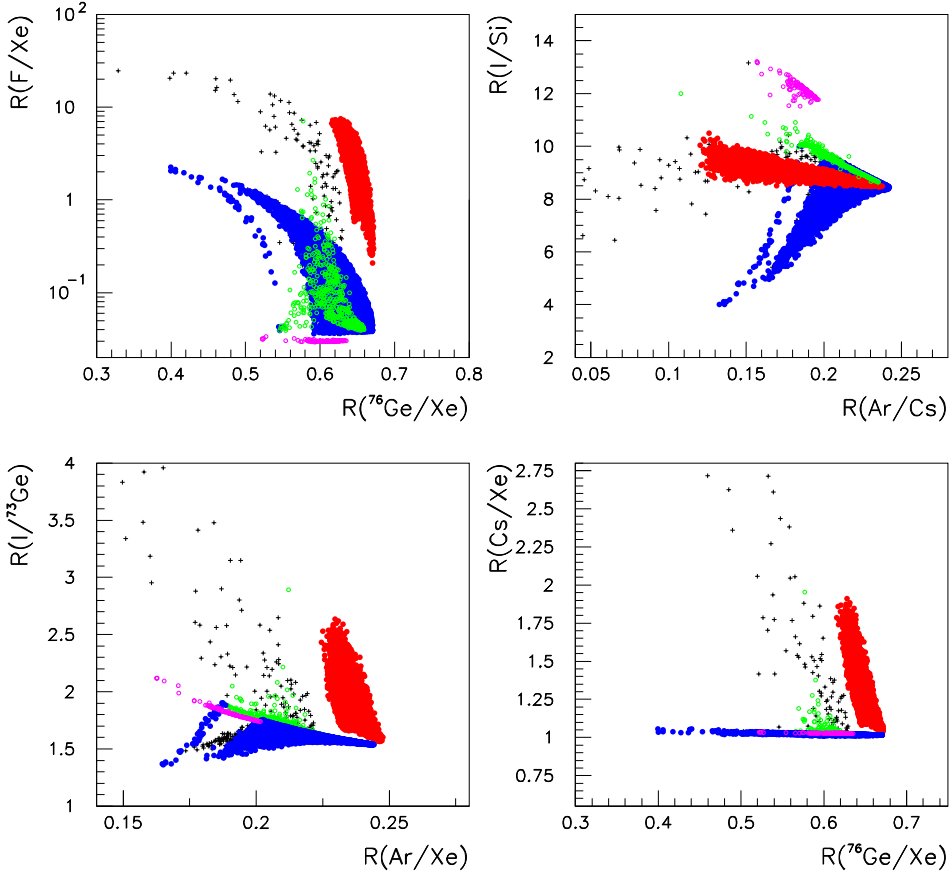,width=14cm}}
\vspace{-1cm}
  \caption{Predictions for the ratio of total rates in 
a)   $R_{F/Xe}$ vs $R_{^{76}Ge/Xe}$, b) $R_{I/Si}$ vs $R_{^{76}Ge/Xe}$, c) $R_{I/^{73}Ge}$  vs $R_{Xe/Ar}$, 
d)   $R_{Cs/Xe}$ vs $R_{^{76}Ge/Xe}$.  Same colour code as fig.~\ref{fig:amp}. }
 \label{fig:nucsi}
\end{figure}

We also considered other combinations of nuclei including those that are primarily sensitive to spin independent  interactions. 
We found that the rates
for  $R_{^{76}Ge/Xe},R_{Ar/Xe},R_{Ar/Cs}$ or $R_{Cs/Xe}$ could also give a  handle to discriminate models.
 The results of our scan of the parameter space in  each of our sample models are displayed in fig.~\ref{fig:nucsi} 
for a representative set of pairs of nuclei. 
As explained before, the rate in $n(Xe)$ increases when $\xi/\lambda_+$ is large and the SD contribution important. This increase is driven by $\xi_n$ so is more important in MSSMH than in MUED or LHM. Therefore reductions in the ratio
$R_{^{76}Ge/Xe}$ or $R_{Ar/Xe}$  are larger in MSSMH than in other models, reaching almost a factor 2. On the other hand  $^{133}Cs$ is sensitive to $\xi_p$ thus the rate $n(Cs)$ can be  large in MUED models leading to a suppression of $R_{Ar/Cs}$  while for other models the predictions for $R_{Ar/Cs}$ are very similar to those for $R_{Ar/Xe}$.  
Note that in the MSSM, there are two disconnected regions in fig.~\ref{fig:nucsi}a-c. Although the  difference is too small to be measured the narrow band corresponds to models with a CDM around 50GeV.

Heavy nucleis can also be used to  identify models where $\tan\phi$ is large. Consider  for example $R_{Cs/Xe}$. 
Both nuclei have similar atomic number and are mainly sensitive to SI interactions, while $^{133}Cs$ is also sensitive to
$\xi_p$ and   $^{129}Xe$ to $\xi_n$.  
In models  where $\xi_p \gg \xi_n$,  such as MUED and some of the MSSMQ and LHM scenarios,  
the ratio $R_{Cs/Xe}$ can be as large as 2 while $R_{Cs/Xe}\approx 1$
in all models where SI interactions dominate and/or $\xi_p\approx \xi_n$.

We conclude that in principle with the observation of signals in detectors with different materials, including detectors highly sensitive to SD
interactions,  discrimination of MUED from MSSMH and LHM is possible except in a small numbers of scenarios. 
On the other hand the model MSSMQ with light squarks can easily  be confused with MUED and LHM.
We do not attempt to estimate the precision to which the ratio of rates can be measured, it is strongly dependent 
on the specific detector in operation. A more precise analysis must await some signals.

\subsection{SI interactions}
\label{sec:sionly}

\begin{figure}
\setlength{\unitlength}{1mm} \centerline{\epsfig{file=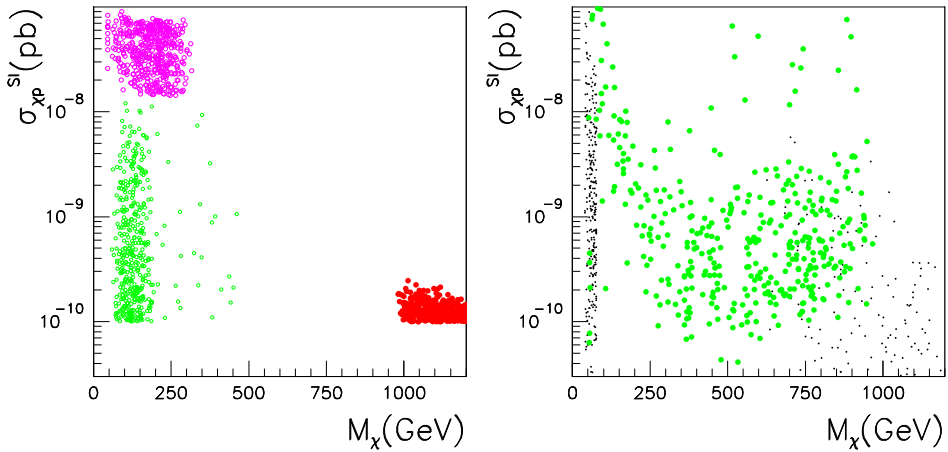,width=14cm}}
\vspace{-1cm}
  \caption{Predictions for $\ssi$ in scenarios where $\ssd<4.\times 10^{-7}$~pb
a)  RHNM (pink), MUED (red), LHM( green) and
b) MSSMH (green) and IDM (black). In b) only models that have $0.094<\Omega h^2<0.136$  are included.}
 \label{fig:sionly}
\end{figure}

The case where only a signal is observed in SI interactions has much less discriminating
power. Basically the only information that can be used is the total cross-section as well
as the mass of the CDM particle.
We have compared predictions for $\ssi$ for all models considered previously 
including only scenarios where a signal would be seen only in the SI channel, that is  $\sigma^{SI}_{\chi N}>10^{-10}$~pb and
  $\sigma^{SD}_{\chi N}<4\times 10^{-7}$~pb. Here we include also the 
IDM which leads only to a signal in SI interactions.
   As discussed previously, the predictions for the RHNM are
 always large $\ssi>1.5\times 10^{-7}$~pb and
 only a light CDM can be expected,
 see fig.~\ref{fig:sionly}a  while those of the LHM are much lower 
$10^{-8}{\rm pb} >\ssi> 10^{-10}$~pb. 
In the MUED model, cross sections
 are very low and the CDM is in the TeV range making it very difficult to see a signal. In the  IDM model
 the CDM is
 either expected to be around 50GeV with cross sections that span over the full range while for heavier CDM particles the
 predictions do not exceed a few $10^{-9}$~pb. 
 In the MSSM (only results for MSSMH are displayed in the figure, similar results are found for MSSMQ)  the whole range of cross sections can be
 expected. For the IDM and MSSM we have imposed the WMAP lower and upper limit for the relic density. 
Allowing other dark matter candidates would lead to many more models passing the constraints and 
cross sections over the whole range for any CDM mass. 
To summarize, with only a signal in the SI channel one could distinguish RHNM
from LHM, while IDM and MSSMH are often indistinguishable from each other and from the previous two models.
A DM mass between $500-700$~GeV is however only compatible with the MSSM. Furthermore a   
heavier DM particle is only compatible with MSSM and IDM models while  no
signal is expected in the MUED model. 
These statements  depend crucially  on the upper bound that can be set on $\ssd$. 
For example, if no signal is observed at the level of $10^{-5}$~pb we would predict that the SI cross 
section in MUED could reach $\ssi\approx 10^{-9}$~pb, an order of magnitude more than what we have used in this section.

\section{Conclusion}

We have summarized the predictions for $\sip$ and $\sdp$ in a variety of new physics models. We  have emphasized the importance of measuring $\sd_N$ to discriminate dark matter models although first signals are generally expected in the SI mode.
This is because  nuclei sensitive to SI interactions  provide a measurement of basically one  specific combination of 
couplings.
On the other hand  both amplitudes $\xi_p$ and $\xi_n$ can be measured with SD sensitive nuclei. 
Furthermore most  particle physics models 
predict $\lambda_p\approx \lambda_n$, at least those where the Higgs is responsible for SI interactions, while $\xi_p/\xi_n$ 
vary over a very wide range, from   $\xi_p/\xi_n\approx -1.1$ in models dominated by Z 
exchange to either very small or very large values when coloured particles play an important role.   
To control astrophysical uncertainties as well as other theoretical uncertainties, we advocate to 
compare ratios of rates measured with different materials. These ratios can be powerful model discriminators 
especially when involving one light nuclei. In particular we have shown that in principle one could disentangle
the MSSMH and MUED models.
Recall that MSSMH is an example of a model with a Majorana neutrino where SI/SD interactions are dominated by H/Z exchange while MUED is a model where the same diagrams (quark exchange) contribute to both type of processes. 
With direct detection alone it is much more difficult to distinguish MUED models from the LHM scenarios  where the new quarks are almost degenerate with
the CDM as well as with  MSSM models with light squarks. This is because in this case the SI/SD amplitudes are also both dominated by the exchange 
of a coloured particle.   
Fortunately if these coloured particles are below the 2~TeV scale, the LHC should be able to detect these new particles.  
 
We have in this analysis concentrated on scenarios with detection rates accessible by future CDM detectors.
There is however no guarantee that a positive signal will be measured for the full parameter space of the models we
have considered. This is especially an issue for light nuclei.  
We have also shown explicitly that if the SD  elastic scattering cross section is below the sensitivity of future detectors, it is much 
more difficult to identify the particle physics model with data from SI direct detection alone, 
 The MSSM, LHM and IDM models all predict $\sip$ in a wide range while a signal should be expected soon in the RHNM. In this case no signal is expected in MUED scenarios.
A determination of the mass of the DM particle will help disentangling some models.

\section{Acknowledgements}
We thank G. Azuelos,  F. Boudjema and V. Zacek for useful discussions.
We thank A. Belyaev for  providing the CalCHEP code for the little Higgs model.
This work is supported in part by the GDRI-ACPP of CNRS and by the French ANR project {\tt ToolsDMColl}, BLAN07-2-194882.
 The work of A.P. was supported by the Russian foundation for Basic Research, grant
RFBR-08-02-00856-a.

\end{document}